\documentclass[letterpaper,12pt]{article}   
\usepackage{osajnl2} 
\usepackage[draft]{hyperref} 
\usepackage{amsmath}
\usepackage{graphicx}
\usepackage{epstopdf}

\begin{document}

\title{Further considerations on layer-oriented adaptive optics for solar telescopes}

\author{Agla\'e Kellerer}
\address{Durham University, \\ 
South Road, Durham, UK\\
a.n.c.kellerer@durham.ac.uk}


\begin{abstract} 

The future generation of telescopes will be equipped with multi-conjugate adaptive optics (MCAO) systems in order to obtain high angular resolution over large fields of view. MCAO comes in two flavors: star- and layer-oriented. Existing solar MCAO systems rely exclusively on the star-oriented approach. Earlier we have suggested a method to implement the layer-oriented approach, and in view of recent concerns we now explain the proposed scheme in further detail. 
We note that in any layer-oriented system one sensor is conjugated to the pupil and the others are conjugated to higher altitudes. For the latter not all the sensing surface is illuminated by the entire field-of-view. 
The successful implementation of nighttime layer-oriented systems shows that the field reduction is no crucial limitation. In the solar approach the field-reduction is directly noticeable because it causes vignetting of the Shack-Hartmann sub-aperture images. It can be accounted for by a suitable adjustment of the algorithms to calculate the local wave-front slopes. 
We dispel a further concern related to the optical layout of a layer-oriented solar system. 
\end{abstract}


\maketitle 

\section{Introduction}
To understand the behavior of the solar magnetic fields, high angular resolution of the solar surface must be attained over large fields-of-view. This calls for MCAO correction on ground-based solar telescopes\,\cite{Collados}. 
In MCAO systems several deformable mirrors are optically conjugated to different turbulent layers in the atmosphere. Each mirror corrects the wavefront distortions introduced close to its conjugate layer. The adaptive control of the mirrors requires the 3D-distribution of the distortions to be reconstructed from a set of wavefront-sensor measurements.

There are two different approaches to achieve the required MCAO wavefront sensing: {\it star-oriented\/} and {\it layer-oriented\/}. Up to now, solar MCAO uses exclusively a procedure that corresponds to the star-oriented approach in so far as each sensor measures the integrated wavefront distortions along one direction. This ``directional'' method entails two difficulties:
\begin{itemize}
\item Adequate inference of the 3D-turbulence from measurements along a few discrete directions is an ill-conditioned problem, and large field-sizes are therefore difficult to correct\,\cite{Berkefeld}. The number of sensing directions can be increased, but eventually the computational load associated with the rapid tomographic reconstruction becomes prohibitive.

\item A Shack-Hartmann (SH) wavefront sensor is employed to determine the integral wavefront distortion in the specified direction. 
In the resulting profiles the high-altitude contribution is most critical for the layer specific adaptive correction in wide fields of view. In the actual nighttime star-oriented method the integration is straightforward, because each lenslet integrates the distortions along a well defined direction. In the solar application of the method each lenslet must form an image of the solar surface with an angular resolution $\sim 0.4''$ which is then correlated against a reference image in order to obtain a local wavefront shift. Since the correlation requires an image sampled over typically $16\times16$ pixels, the viewing field of a lenslet must have an opening angle of $\sim 6.4''$. For adequate resolution the lenslet needs to have a diameter $\sim 0.10$\,m at the telescope pupil. The sub-aperture diameter will then be $\sim 0.41$\,m at altitude 10\,km. Sampling regions at the critical high altitude that overlap substantially and are much larger than at ground level are an undesirable attribute of the directional, star-oriented method. 
\end{itemize}

As an alternative to the star-oriented approach we have proposed a layer-oriented set-up for solar MCAO\,\cite{LO}. 
In the star-oriented approach the lenslets of the SH sensors are conjugated to the telescope pupil.
In the layer-oriented approach they are conjugated to a number of turbulent layers above the pupil, and each deformable mirror is paired with a SH sensor conjugated to the same altitude. This has several advantages:

\begin{itemize}
\item For a SH sensor conjugated to a layer at altitude $h$, the effective sub-aperture size is smallest at this altitude $h$ and increases with distance from that layer. The fluctuations are thus determined with best resolution near the layer of interest, the contributions from other layers are attenuated, i.e. are averaged out over larger sub-apertures. This reduction is largest for the large fields of view required in solar observations. While the directional, star-oriented approach is limited to fairly narrow fields of view, the layer-oriented approach works best with wide fields of view. 

\item Each SH sensor images the entire science field. The sensor measurements for the AO correction cover, thus, the entire field, while in the star-oriented approach the correction needs to be extrapolated from measurements along a few discrete directions. 

\item The cross-correlation is done over large fields, e.g. over $100\times100$ pixels for a $40''$ field sampled at $0.4''$. The quality of the cross-correlation is thus far better than with $16\times16$ pixels in the star-oriented approach. 
\end{itemize}
The main difficulty of the layer-oriented approach is the need for fast detectors with a large number of pixels. 

\section{Recent concerns on the layer-oriented method}

Analyzing the layer-oriented MCAO approach for solar observations Marino \& Woger\,\cite{MW} have suggested that the method is not viable due to vignetting. 
We appreciate their effort and take the opportunity to give further details on the layer-oriented method. 

A feature inherent to any layer-oriented MCAO approach is that not all elements of sensors conjugated above the telescope pupil (to a so-called {\it meta-pupil\/}) are illuminated by the entire field-of-view. 
Nighttime layer-oriented systems employ pyramid sensors\,\cite{Ragazzoni}. A few pyramids average the wavefront shift over  the reduced field of view.  Because pyramid sensors produce pupil images rather than field images, the field reduction does not become visible as vignetting, but it weakens equally the signal attenuation from unconjugated layers. 
The successful use of pyramid-based systems shows that the field reduction is no crucial limitation.  We discuss this in Section\,\ref{sec:1}.

In the solar approach the lenslets of the SH wavefront sensor produce field images which are then intercorrelated to determine the average phase shifts over the field\,\cite{LO}. The field reduction becomes apparent as vignetting in these images. The required adjustments of the existing data-analysis techniques are discussed in Section\,\ref{sec:2}. 

As a further concern about the feasibility of the set-up it has been suggested that large fields require impractically small focal lengths\,\cite{MW}. Section\,\ref{sec:3} clarifies that this conclusion relies on the (unnecessary) assumption of a fixed {\it pitch\/}, i.e. size, of the SH lenslets. 
In fact, there is no need to use lenslet arrays with the same pitch for different conjugate layers or for different field-sizes. 
If the pitch is adjusted, the focal length can be sufficiently large.

\subsection{Field reduction}\label{sec:1}

\begin{figure}[htbp]
\begin{center}
\includegraphics[width=.5\textwidth]{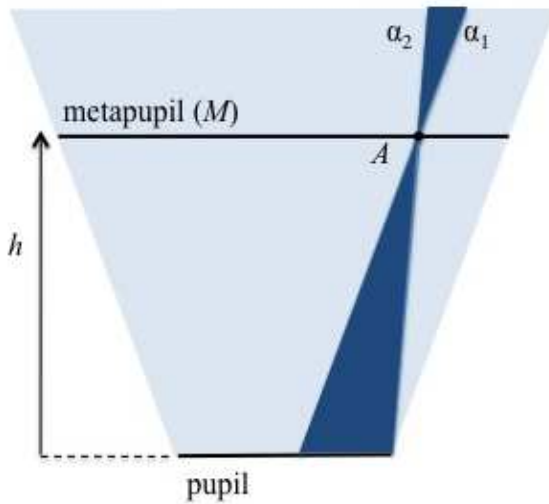}
\caption{Light blue region: field-of-view of the science image and of the wavefront sensors. A SH-lenslet conjugated to the position $A$ estimates the wavefront distortion within the reduced field $\alpha_1$ to $\alpha_2$ (dark blue region). The field reduction is not a feature specific to the solar layer-oriented approach, it affects pyramid-based systems in the same way as SH-based systems; the difference is merely that the vignetting becomes visible in the SH-images. }
\label{fig:1}
\end{center}
\end{figure}

Any MCAO system that aims at compensating atmospheric distortions introduced over a particular science field needs a set of sensors that cover that field. The blue area in Fig.\,\ref{fig:1} indicates the science field over which MCAO correction is required. Only the atmospheric turbulence within this region contributes to the image degradation within the chosen science field. 

The horizontal plane $(M)$ that intersects the blue region in Fig.\,\ref{fig:1} is the meta-pupil at altitude $h$. Whatever sensor is used -- pyramid, SH or other -- a sensor element conjugated to the point $A$ on the meta-pupil experiences the wavefront distortions within the double-cone $\alpha_1$ to $\alpha_2$. Only this reduced field (deep blue region in Fig.\,\ref{fig:1}) determines the distortion sensed at point $A$. This is a feature of any layer-oriented MCAO, it  affects pyramid-based systems in the same way as it affects SH sensors.

The attenuation of the contribution from unconjugated layers is substantially stronger on a solar image than on a nighttime image. This is so because the averaging relates to a continuous field, rather than a discrete number of directions. In nighttime systems, the number of directions is restricted by the number of guide-stars within the field, and typically it lies between 3 and 8\,\cite{Arcidiacono}. 
Nevertheless, nighttime layer-oriented adaptive-optics functions\,\cite{Arcidiacono}. Since the field-reduction in high-altitude conjugation affects solar systems equally, it will likewise not impede their performance.

\subsection{Vignetting}\label{sec:2}

A feature specific to the use of SH wavefront sensors is the vignetting  of the images behind the lenslets.  
This affects the cross-correlation and it has been argued\,\cite{MW} that ``this is particularly problematic for sub-apertures on opposite edges, which share no common un-vignetted field points. These effects make the successful cross-correlation of sub-aperture images a practical impossibility.''

Vignetting does complicate the correlation of two images, i.e. it requires some modification of the correlation algorithm. However, correlation remains possible for images with substantial overlap, and this is particularly so for a large field width because the images contain then large pixel numbers, such as $200\times200$ for $\alpha= 80''$. 
 When two images do not contain sufficient overlap they can be correlated indirectly through other images. In particular there is no need to cross-correlate sub-aperture images recorded on opposite edges of the pupil.
Three familiar approaches can exemplify the cross-correlation of vignetted sub-aperture images:

\begin{enumerate}
\item If the central sub-aperture image is complete it can  be correlated against all other sub-aperture images. A masking function on the reference image accounts for their vignetting; it equals 0 where the field direction is entirely vignetted, and 1 where it is not. 
The central image  remains complete up to height $h=(D-d)/\alpha$, where is the sub-aperture diameter, $D$ the telescope aperture, and $\alpha$ the field width.
For the next generation of 4\,m class solar telescopes, $h= 9.5$\,km, if $d=0.3$\,m and $\alpha=80''$. 
For a current 1.6\,m solar telescope and the same field width the central image begins to be incomplete at 3.6\,km, if $d=0.2$\,m. Once the central image is reduced, the correlation can be step-wise, i.e. the outer images can be correlated to intermediate images that are correlated to the central image.

\item If the SH wavefront sensors have the same field-of-view and pixel scale, one can use an image of the ground-layer sensor as reference for the high-altitude sensors. This relies on the alignment between sensors which is required at any rate for successful MCAO correction. In this approach one applies again a direction-dependent masking function that accounts for vignetting. Note that the reference images are updated only every $\sim 20$ seconds, so that -- if the different sensors are connected to different computers -- the data from the reference image can readily be transferred between computers.

\item An alternative to the inter-correlation of images from different sub-apertures is cross-correlation of each sub-aperture against itself, i.e. correlation of an image recorded at time $t$ against one at time $t_0$. No masking function is then required, one compares the shift at time $t$ against the running average of shifts. 
\end{enumerate}

As shown in sections\,\ref{sec:1} and \ref{sec:2}, field-reduction is a feature of layer oriented MCAO systems. 
The vignetting it induces in the SH images can be accounted for by suitable modifications of the correlation algorithms.  
The main difficulty appears to be the need for very large and fast detectors in order to achieve the essential advantages of the method: the use of the full field information and the improved sensitivity to high-altitude turbulence.

\subsection{Consideration of the optical set-up}\label{sec:3}

A further concern that can be addressed is the claim that -- in a layer-oriented system -- large fields require lenslet arrays with impractically small focal lengths\,\cite{MW}. 
A lenslet array of pitch $p$ and focal length $f$ can image a maximum angular diameter $p/f$. 
Each lenslet spans a sub-aperture of size $D_{\rm m\/}/N$ in the meta-pupil, where $D_{\rm m\/}=D+\alpha\,h$ is the diameter of the meta-pupil and $N$ equals the number of sub-apertures across the beam diameter. The on-sky field-diameter, $\alpha$, is then obtained via the Lagrange invariant: 
\begin{eqnarray}
\alpha \,\cdot \frac{D_{\rm m\/}}{N} &=& \frac{p}{f} \cdot p \\
\alpha &=& \frac{p^2\,N}{f \, D_{\rm m\/}}
\label{eq:alpha}
\end{eqnarray}

From this relation it has been concluded that, for a given pitch, short focal lengths are required to image large fields of view\,\cite{MW}. 
However there is no need for sensors conjugated to different layers to have the same lenslet pitch.    Let $s$ be the size of the detector pixels. The cross-correlation of solar images requires an angular sampling $\theta\leq 0.5''$. The pitch is related to the field diameter through the relation:
$p = s \,\alpha/\theta$. 
For an angular sampling $\theta=0.4''$ and a pixel size $s=7\,\mu$m, the pitch increases from 0.7\,mm to 3.5\,mm for field diameters $\alpha=40''$ to $200''$. As it happens a large pitch facilitates system alignment and thereby improves image quality.

The number of sub-apertures across the beam-diameter equals $N=D_{\rm m\/}/d$, where $d$ is the sub-aperture diameter. 
Hence, Eq.\,\ref{eq:alpha} is rewritten as:
\begin{eqnarray}
f = \frac{s^2 \, \alpha}{d\, \theta^2}
\end{eqnarray}
For a given pixel size, $s$, the focal length of the lenslets increases with the field diameter. There is thus no basis for the objection that large field sizes require lenslets with impractically small focal lengths. 

\begin{figure}[htbp]
\begin{center}
\includegraphics[width=.35\textwidth]{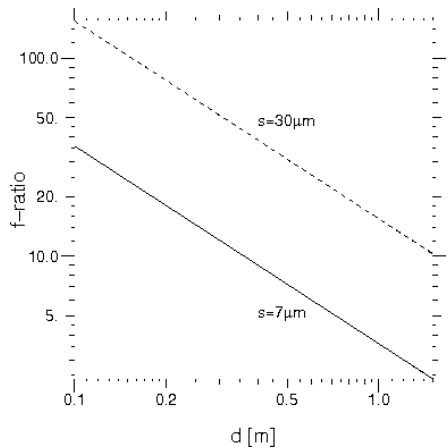}
\caption{The $f$-ratio of a SH lenslet as a function of the sub-aperture diameter, $d$, for $0.4''$ pixel sampling and a pixel size $s$. }
\label{fig:fratio}
\end{center}
\end{figure}

Fig.\,\ref{fig:fratio} traces the $f$-ratio, $f/p$, as a function of sub-aperture diameter, $d$, for an angular sampling $\theta=0.4''$ and for pixel sizes $s=7\,\mu$m and $30\,\mu$m. 
The sub-aperture diameter of a sensor conjugated to height $h$ should roughly equal the Fried parameter of the turbulent layer at altitude $h$.
For a sensor conjugated to the ground layer $d\sim 0.1$\,m, while for weaker high-altitude layers: $d \geq 0.5$\,m. 
As seen from Fig.\,\ref{fig:fratio}, the $f$-ratio decreases with sub-aperture diameter, $d$, but can be kept large by increasing the pixel size.

\section{Conclusion}

In a layer-oriented adaptive optical system a deformable mirror and a wavefront sensor are conjugated to a dominant turbulent layer. A complication of this method is that, for sensors conjugated to the high altitudes, the sensing surface is illuminated by only part of the field-of-view. This field reduction is inherent to any layer-oriented approach. It weakens the attenuation, i.e. the averaging out of the fluctuations in distant layers. However, in the solar application the signal is averaged over a continuous field, rather than a limited number of directions, and the attenuation of distant layers remains therefore considerably more efficient than in the pyramid-based nighttime systems\,\cite{Ragazzoni, Arcidiacono}. Since these latter systems are successfully used on-sky, the field reduction will be even less critical than in the solar application. Compared to the directional, `star oriented' method, the essential advantage of the layer oriented method is that it focusses on -- and thereby achieves optimal resolution at -- the conjugated layers.

A second feature of the solar layer-oriented method is that the field reduction causes the images behind the SH sensors to be vignetted. The vignetting requires an adjustment of the correlation algorithm to determine the local wavefront slopes, but this adjustment is straightforward in view of the broader field images that are attained with the method.

Under the implicit assumption of a constant lenslet pitch it has been claimed that  lenslet arrays with excessively small focal lengths are needed on large fields-of-view\,\cite{MW}. However, if detector pixels of  constant size are used, the focal length of the lenslets increases with field size and the lenslet pitch increases likewise. As it happens, this facilitates system alignment and thereby improves image quality. 

In conclusion, field-reduction is inherent to any layer-oriented approach and is no critical limitation. The specific complication of vignetted SH images can be accounted for by standard methods.  The main difficulty of the new layer oriented multi-conjugate adaptive optical system is the need for fast detectors with a large number of pixels. While this is a limitation today, technological advances will resolve it before long. 

\bibliographystyle{unsrt}
\bibliography{References}

\end{document}